# Complexity-weighted doses reduce biological uncertainty in proton radiotherapy planning

Stephen J McMahon PhD[1] *, Harald Paganetti PhD[2], Kevin M Prise PhD[1]

1 Centre for Cancer Research and Cell Biology, Queen's University Belfast, Belfast, Northern Ireland

2 Department of Radiation Oncology, Massachusetts General Hospital and Harvard Medical School, Boston, MA, USA

* Corresponding author: Stephen.mcmahon@qub.ac.uk; Tel: +44 28 9097 2620

# Abstract

*Purpose*

Variations in proton Relative Biological Effectiveness (RBE) with Linear Energy Transfer (LET) remain one of the largest sources of uncertainty in proton radiotherapy. This work seeks to identify physics-based metrics which can be applied to reduce this biological uncertainty.

*Materials and Methods*

Three different physical metrics – dose, dose × LET and a complexity-weighted dose (CWD, $Dose \times (1 + \kappa\, LET_D)$) ) were compared with *in vitro* experimental studies of proton RBE and clinical treatment plans analysed using RBE models. The biological effects of protons in each system were plotted against these metrics to quantify the degree of biological uncertainty introduced by RBE variations in each case.

*Results*

When the biological effects of protons were plotted against dose alone, significant biological uncertainty was introduced as the LET-dependence of RBE was neglected. Plotting biological effects against dose × LET significantly over-estimated the impact of LET on cell survival, leading to similar or greater levels of biological uncertainty. CWD, by contrast, significantly reduced biological uncertainties in both experiments and clinical plans. For prostate and medulloblastoma treatment plans, biological uncertainties were reduced from ±5% to less than 1%.

*Conclusions*

While not a replacement for full RBE models, physics-based metrics such as CWD have the potential to significantly reduce the uncertainties in proton planning which result from variations in RBE. These metrics may be used to identify regions in normal tissues which may see unexpectedly high effects due to end-of-range elevations of RBE, or as a tool in optimisation to deliver uniform biological effects.

## Introduction

Proton radiotherapy offers significant dosimetric advantages over conventional photon-based therapies, as proton's continuous energy loss gives them a well-defined range with greatest energy deposition at the end of their path. As a result, normal tissue doses can be significantly reduced, reducing side-effects or allowing an escalation of target doses [1].

However, using protons is not problem-free. Protons bring new challenges in dosimetry and planning, as uncertainties in the tissue composition and particle range can lead to classes of uncertainty not seen in photon radiotherapy [2,3]. But perhaps more significantly, there are also differences in biological effectiveness between photons and protons which are not typically incorporated in clinical planning.

When different particles are used to deliver the same dose of ionising radiation, differences in biological effects are observed, typically defined in terms of a Relative Biological Effectiveness (RBE) – the ratio between the dose needed to produce a particular biological response with the test radiation and the dose of a reference radiation which produces the same effect. For protons, a constant RBE of 1.1 is typically assumed, and plans are optimised based on physical dose alone [4].

While this reflects that protons are more biologically effective than photons, it assumes that this effect is a constant independent of dose, tissue type and LET, despite extensive preclinical evidence that proton RBEs are variable. This reflects a significant limitation in proton treatment planning, as it introduces large biological uncertainties into planning, which can be many times larger than acceptable limits for physical uncertainty. As a result, proton therapy is not delivering on its full potential. Incorporation of the LET-dependence of proton RBE could reduce treatment uncertainties and potentially identify RBE 'hotspots' which may drive unexpectedly large responses, particularly at the end of range which typically occurs in normal tissue [5].

Empirical analytic proton RBE models have been developed to better understand and quantify these effects. Typically, these models begin from the linear-quadratic cell survival model, and adjust the α and β parameters with the introduction of LET-dependent terms [6–8]. While these models broadly agree on overall trends in proton RBE for cell survival, they have seen limited translation to the clinic. In part, this is because these models depend on both the α/β ratio of the target tissue as well as model-specific fitting parameters which carry significant uncertainties and potential inter-patient variability, which translates into significant uncertainties in predicted RBE [9–11]. Furthermore, these models are based on cell survival endpoints that may not be relevant for normal tissue toxicities.

But it is important to note that even in photon-based therapy treatments are planned based on physical dose rather than biological effect, and there is considerable biological uncertainty around the effects of a given dose – both between different tissues and between different patients [12–14]. Despite this, dose-based planning is effective because tissue response is monotonic with dose: increasing dose smoothly and continuously increases effect, even if the exact level of cell killing is unknown.

Similar ideas can be applied in proton therapy – even if there remains significant uncertainties in RBE, physically-determined metrics which correlate better with response have the potential to serve as useful tools for treatment optimisation. In this vein, groups have proposed optimisations based on the physical quantities of dose and LET as a proxy for response, rather than RBE models [15–18]. However, it remains unclear how effectively these approaches reduce biological uncertainties, or what metric is best suited to this purpose. In this work, we evaluate dose, dose × LET and a weighted average of these quantities to evaluate how effectively they can be used to reduce RBE-related biological uncertainties.

## Methods

*Radiation Response Metrics and Biological Uncertainty*

When cells are exposed to a given proton dose, a range of survivals may be observed due to the LET-RBE relationship whose biology is not described by dose alone. In this work, this range of responses seen for a given dose (or other response metric) is referred to as the 'biological uncertainty'.

Three response metrics have been studied: physical dose, $D$; dose-LET product, $D \times LET_D$; and a weighted average of these metrics, called Complexity Weighted Dose (CWD) given by $D_{CW} = D + \kappa\, D \times LET_D = D \times (1 + \kappa\, LET_D)$, where $\kappa$ is an empirical fitting parameter with units of µm/keV. While this approach can potentially be applied to either track- or dose-averaged LET, within the work we have focused on dose-averaged LET ($LET_D$).

*Cell Survival Data*

Survival data for cells exposed to protons was obtained from previous publications for AGO-1522 fibroblasts and U87 glioma cells [19,20]. Clonogenic cell survival was measured at various positions across either pristine or spread-out Bragg peaks (SOBPs) to quantify the dose response curves at different energies, with doses from 0 to 7 Gy and $LET_D$ from 0 to 25 keV/µm. For each position, $LET_D$ was calculated using Monte Carlo simulations of the beamline using Geant4 [21]. Further details on experimental design can be found in the original papers [19,20].

Survival from these experiments was then plotted against each of the three candidate metrics. Correlation between each metric and biological response was assessed by fitting a function of the form $\ln(S) = -AM - BM^2$, where $M$ is one of the three metrics, and $A$ and $B$ are fitting parameters.

For $M = D$, this reduces to the standard linear-quadratic dose response model with $A$ and $B$ taking units of Gy$^{-1}$ and Gy$^{-2}$. These units are also used for complexity weighted dose. For the $M = D\ LET_D$ case, A and B take units of µm keV$^{-1}$ Gy$^{-1}$ and um² keV$^{-2}$ Gy$^{-2}$, respectively. While the dose-LET product can be rescaled to units of Gy by multiplying by a $\kappa$-like term, this does not impact on the overall correlation and so was not done.

*Spread Out Bragg Peak RBE Comparison*

Experimental data necessarily covers only a limited range of combinations of dose and $LET_D$. To evaluate the metrics at a wider range of conditions and to better understand their underlying behaviour in this experimental dataset, RBE was modelled analytically across the whole SOBP. Dose and $LET_D$ distributions were calculated for a 2 Gy SOBP delivered with 62 MeV protons with 10 mm range modulation, corresponding to that used experimentally [19].

At each point along the SOBP, RBE-weighted doses were calculated based on experimentally-determined parameters, with survival given by:

$$S = e^{-(\alpha_x + \lambda LET_D)D - \beta_x D^2}$$

Where $\alpha_x$ and $\beta_x$ are the experimentally-determined X-ray radiosensitivity parameters, and $\lambda$ is an empirical fitting parameter describing the LET dependence of $\alpha$ for protons. For these experiments [19,20], these parameters were determined to be: $\alpha = 0.77 \pm 0.02$ Gy$^{-1}$, $\beta = 0.06 \pm 0.01$ Gy$^{-2}$ and $\lambda = 0.06 \pm 0.01$ Gy$^{-1}$ µm keV$^{-1}$ for AGO-1522 cells; and $\alpha = 0.13 \pm 0.02$ Gy$^{-1}$, $\beta = 0.05 \pm 0.02$ Gy$^{-2}$ and $\lambda = 0.026 \pm 0.003$ Gy$^{-1}$ µm keV$^{-1}$ for U87 cells. These RBE-weighted doses were then plotted against each of the three metrics.

*Clinical Treatment Plan Comparison*

Finally, two realistic clinical treatment plans were evaluated: one medulloblastoma and one prostate. Clinically delivered treatment plans were re-simulated using TOPAS [22] as described elsewhere [8] to provide dose and $LET_D$ distributions for the plans.

As specific survival data is not available for the tissues in these treatment plans, RBEs were calculated using a previously published phenomenological model [8]. This model includes tissue α/β ratios and four empirical model parameters which were fit to published experimental data. RBE-weighted doses were then calculated for each plan, with α/β values set to 10 for the medulloblastoma, 1.5 for the prostate, and 3 for all normal tissues. RBE-weighted doses were then compared to each of the three response metrics for all voxels seeing at least 1% of the prescription dose.

*Statistical Analysis*

A single value of $\kappa$ was used throughout this work for all endpoints and cell types. This value was obtained by carrying out least-squares fitting to the experimental survival data in both cell lines as presented in Figure 1 c and f.

# Results

*Cell Survival*

Experimental cell survival values are plotted against each of the response metrics in Figure 1 for AGO-1522 and U87 cells. Plotting against dose alone (a, d) shows the well-established LET dependence of proton RBE, with an average fit over-estimating cell killing at low LET and under-estimating cell killing at high LET. This leads to a significant biological uncertainty at all doses.

When considering $D \times LET_D$ (b, e), this trend is reversed, with the average fit under-estimating killing at low LET, and over-estimating at high LET, with a significantly greater biological uncertainty than seen for dose alone. This highlights that $D \times LET_D$ alone is not a useful metric for biological response, but the trend is in line with suggestions that it is a proxy for additional biological effect, suggesting that a combined metric may be useful.

This combined metric, the complexity-weighted dose, $D_{CW} = D \times (1 + \kappa\, LET_D)$, is plotted in the lower panels (c,f), with a fitted $\kappa = 0.055 \pm 0.003$, showing significantly improved correlation. $R^2$ values were 0.84 and 0.88 for AGO-1522 and U87 cells when fit using dose alone, increasing to 0.99 and 0.95 when fit using CWD. Applying CWD with a single κ parameter thus accounts for 60 to 90% of the LET-related uncertainty in survival in these cells, despite the different overall radiosensitivities and X-ray $\alpha/\beta$ ratios of these cell lines (12 Gy and 2.5 Gy).

*Spread Out Bragg Peak RBE*

Figure 2 presents dose and $LET_D$ distributions across a 2 Gy SOBP delivered using 62 MeV protons with a 10 mm range modulation, together with RBE-weighted doses for AGO-1522 and U87 cells using empirical fitting to the experimental data in Figure 1.

For each position along the depth-dose curve, RBE-weighted doses were compared to the different dose metrics, shown in Figure 3. Comparing dose alone to RBE-weighted dose shows similar trends to survival, under-estimating the effects of high LET exposures (a, d). In these plots, trends across the

SOBP can be observed – initially, there is good correlation with dose in the low-LET entrance region (dark points) where RBE variations are small. But within the SOBP, dose is roughly constant while LET and RBE increase significantly. And then in the distal tail, dose falls steeply while RBE continues to increase, giving rise to a wide spread in possible biological effects at a given dose.

A similar spread is seen comparing with $D \times LET_D$, although once again the trends with LET are reversed (b, e). An initially rapid rise is seen as dose increases, but within the SOBP the slope is much shallower, as the dependence of RBE-weighted dose on LET is much less than its dependence on physical dose. This also leads to a significant over-estimation of effect in the distal tail.

Applying CWD using $\kappa = 0.055$ dramatically reduces biological uncertainties (c, f), as both physical dose and LET are appropriately taken into account. In both cases uncertainties are substantially reduced, with overall performance slightly better in AGO-1522 than in U87 cells. It is important to note that these plots show some curvature rather than the direct proportionality which would be expected of a 'true' RBE model, as CWD deliberately does not seek to take into account differences in the underlying tissue biology. However, by providing a monotonic correlation with effect, it can effectively identify regions of high biological effect for use in treatment design or optimisation.

*Clinical Treatment Plans*

Finally, these effects have been calculated for medulloblastoma and prostate clinical treatment plans, with dose and $LET_D$ distributions shown in Figure 4. In common with most proton treatment plans, these approaches have relatively low LETs in the entrance region and much of the target volume, but higher LETs at the end of range and at edges of treatment fields. For each voxel in the treatment plan, an RBE-weighted dose has been calculated using the model of McNamara et al [8] and compared against each of the three response metrics, plotted in Figure 5.

Comparing dose alone (a, d) to RBE-weighted dose illustrates the range of biological uncertainty in these plans if they are optimised on physical dose as is currently the case. There is significant variation across the whole dose range, equivalent to ±5% of the prescription dose or more. This is again due to the lack of incorporation of RBE variability, with higher LET regions seeing increased biological effects. $D \times LET_D$ (b, e) is ineffective at resolving this uncertainty, over-estimating the impact of LET and introducing new biological uncertainties.

However, the weighted dose metric (c, f) greatly reduces this biological uncertainty. By applying the same $\kappa = 0.055$ as in the experimental dataset, biological uncertainties can be reduced to less than 1% across the whole dose range. This suggests that this approach may effectively help to identify regions of high biological effect in treatment planning and optimisation.

*Sensitivity to κ*

A single value of $\kappa = 0.055$ has been used throughout this work, determined by minimising the biological uncertainties in an initial *in vitro* experimental dataset. To evaluate the sensitivity of this model parameter, we calculated the average biological uncertainty across the two clinical plans for a range of values of κ and $\alpha/\beta$. Biological uncertainty was defined as the 95% confidence interval of biological effect, plotted in Figure 6.

In the limit of $\kappa = 0$, this gives the biological uncertainty according to dose alone, which is approximately ±5% for a normal tissue α/β of 3 Gy as shown in Figure 5. As κ increases, these uncertainties decrease, reaching a minimum of less than 1% around $\kappa = 0.055$. This minimum is relatively shallow, with κ values from 0.04 to 0.08 all yielding average biological uncertainties less than 1.5%, suggesting that this metric still offers good improvements in predictive power even if the κ value is not precisely optimised. A shaded range is also plotted showing the range of biological uncertainties for $\alpha/\beta$ ratios from 2 to 10 Gy. While the exact degree of uncertainty varies considerably depending on the tissue $\alpha/\beta$, $\kappa = 0.055$ also provides a minimisation of the worst-case error.

## Discussion

Fully optimising treatment plans remains a significant challenge in radiotherapy, particularly in proton therapy where the LET dependence of biological response adds extra potential dimensions of optimisation [23]. However, despite extensive evidence that proton RBE is variable, most clinical approaches adopt the conservative approach of a constant RBE of 1.1 [3].

In this work, we explored the use of physics-based metrics to reduce the biological uncertainties in radiation response. In particular, we sought to identify metrics which were monotonically related to survival, without explicitly calculating RBE values. By not attempting to generate equivalent photon doses, many aspects of the underlying biology which are subject to large uncertainties (such underlying tissue radiosensitivity and α/β ratio) can be neglected, greatly simplifying this calculation. Despite this simplification, we find that CWD significantly reduces uncertainty in level of response when compared to dose or $Dose \times LET_D$ alone, reducing uncertainty from ±5% to ±1% in clinical treatment plans.

Such an approach has significant potential for applications in proton therapy. A major ongoing area of research is in optimising therapy to account for elevated end-of-range LETs in normal tissue which may lead to increased complication rates [24]. This metric can help guide such studies by identifying regions would be expected to show the greatest deviation from physical dose alone. In addition, while optimising dose and LET independently may go some way to addressing RBE effects [17,18], because of the interplay between these parameters it is possible that reduction in physical dose will be offset by increases in LET, or vice-versa. By reducing this uncertainty, optimisation of complexity-weighted dose has the potential to offer more biologically robust solutions [3].

It is important to note that this approach is not a true RBE model. The relationship between CWD and RBE-weighted dose (Figures 3 & 5), because variations in tissue sensitivity and α/β are not taken into account. This also means that these curves will be different for different tissues – as can be seen in Figure 5, where separate small clusters of points are seen at high doses, corresponding to tumour volumes with different α/β ratios to normal tissue (a higher α/β value leading to lower RBEs in the medulloblastoma case, and vice versa for prostate). However correlations remain good within a given tissue, enabling robust optimisation of effects as in photon therapy.

Finally, this work, a value of $\kappa = 0.055$ has been used based on *in vitro* experimental data, which was seen to perform well across the range of scenarios considered here and may already be an effective tool in optimising therapy. However, as this approach is necessarily an approximation to the true underlying biology, the true optimum value will depend on the chosen target, tissue α/β ratios, and the dose and $LET_D$ characteristics of the treatment plan. More comprehensive planning and analysis studies are needed to fully assess the performance of this model in a range of scenarios and what impact these choices may have on the clinic.

## Conclusions

A complexity-weighted dose (CWD) metric, $D_{CW} = D \times (1 + \kappa\, LET_D)$, has been found to effectively reduce biological uncertainties resulting from variable RBE in proton radiotherapy. While this cannot be used to set absolute dose constraints, it effectively identifies regions which may see elevated biological effects due to high LET, and may prove a valuable tool for therapy optimisation and retrospective response analysis.

## Acknowledgements

SJM would like to thank the European Commission (EC FP7 grant MC-IOF-623630) for supporting this work. The authors would like to thank Aimee McNamara and Drosoula Giantsoudi for their assistance in preparing the clinical treatment plans.

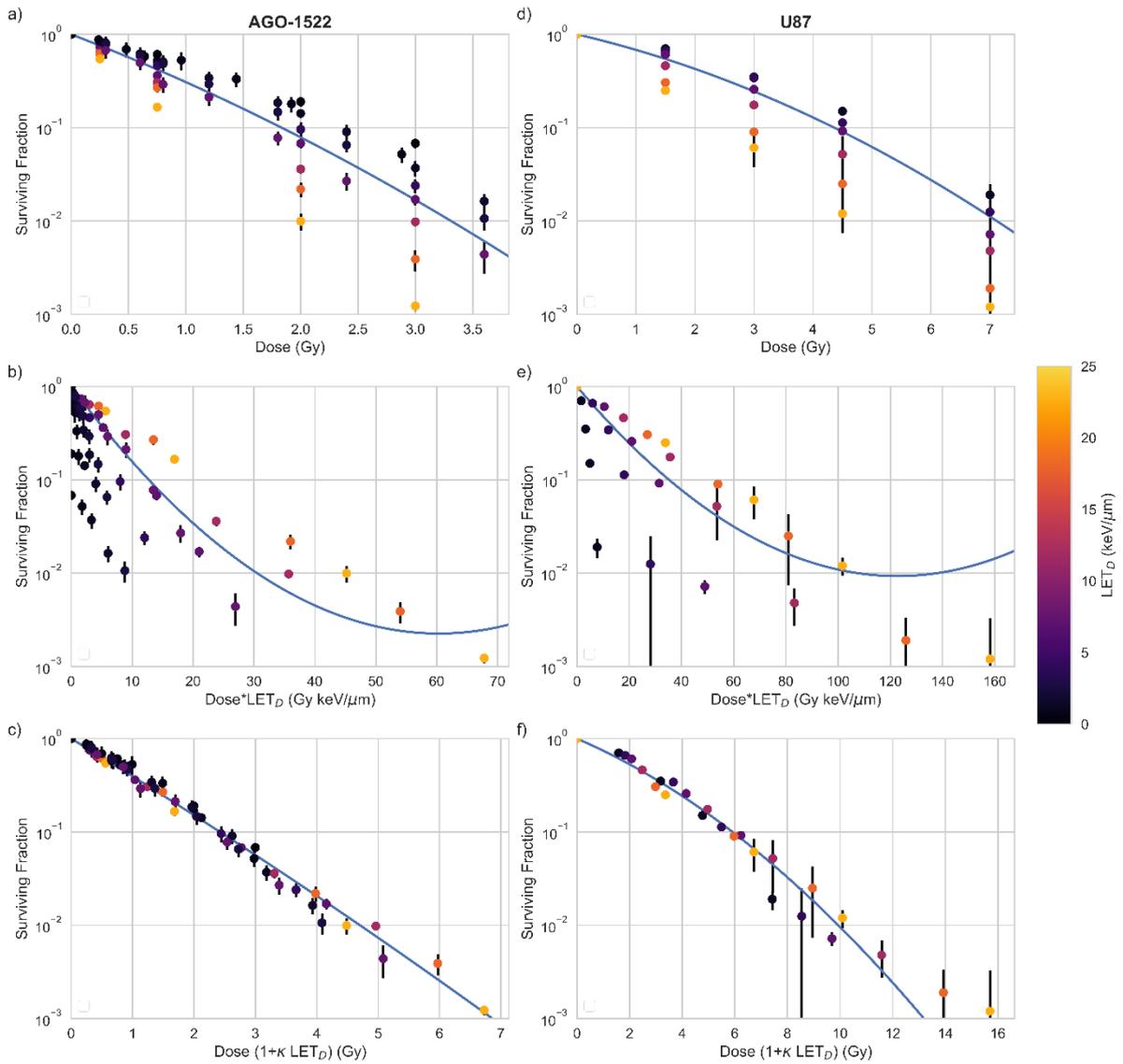

Figure 1: Uncertainty in cell survival compared to different response metrics for AGO-1522 fibroblast (left) and U87 glioma (right) cells[19,20]. Each point represents a single condition, coloured according to LET. When survival is plotted against dose (top), the LET-dependence of proton RBE is seen, with increasing cell kill at higher LET. Plotting against $D \times LET_D$ (middle) sees greater biological uncertainty, although with a reversed LET dependence. Plotting against CWD (bottom), however, gives excellent correlation across all LETs, greatly reducing biological uncertainties.

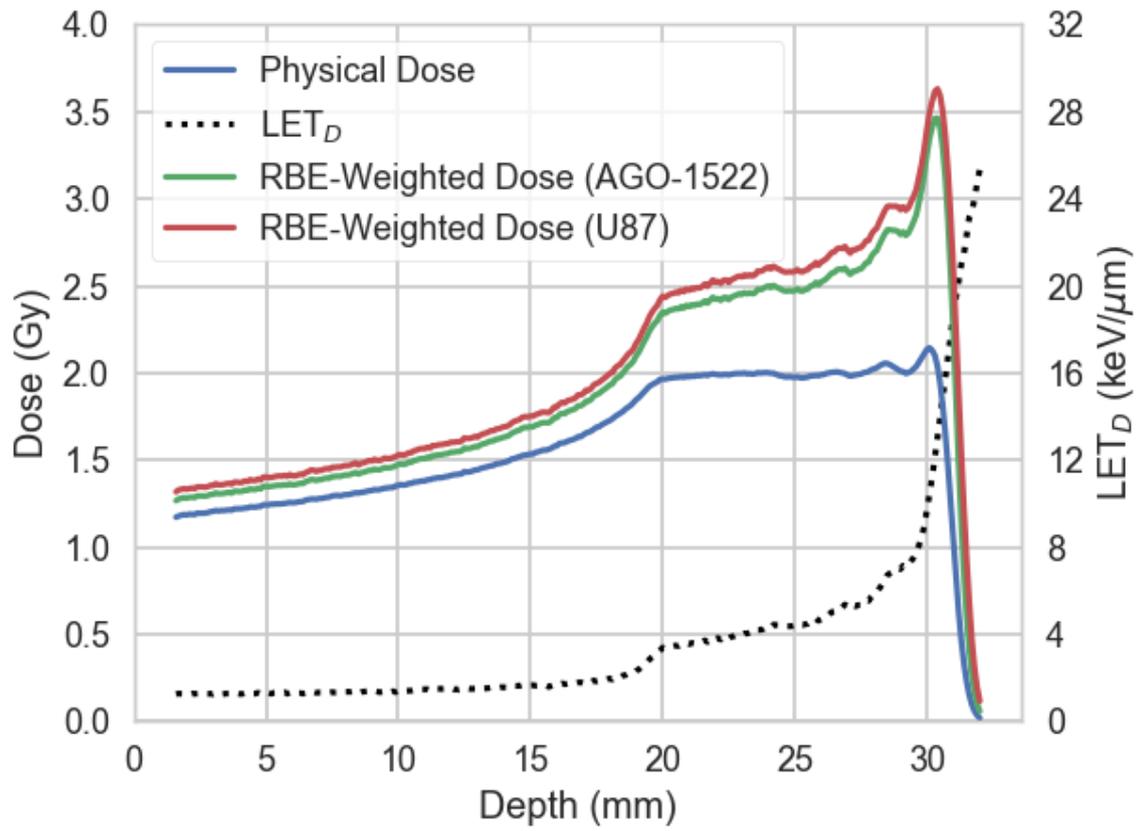

Figure 2: Dose and $LET_D$ across 62 MeV SOBP with 10 mm modulation, along with RBE-weighted doses for two cell lines based on empirical fitting to data in Figure 1.

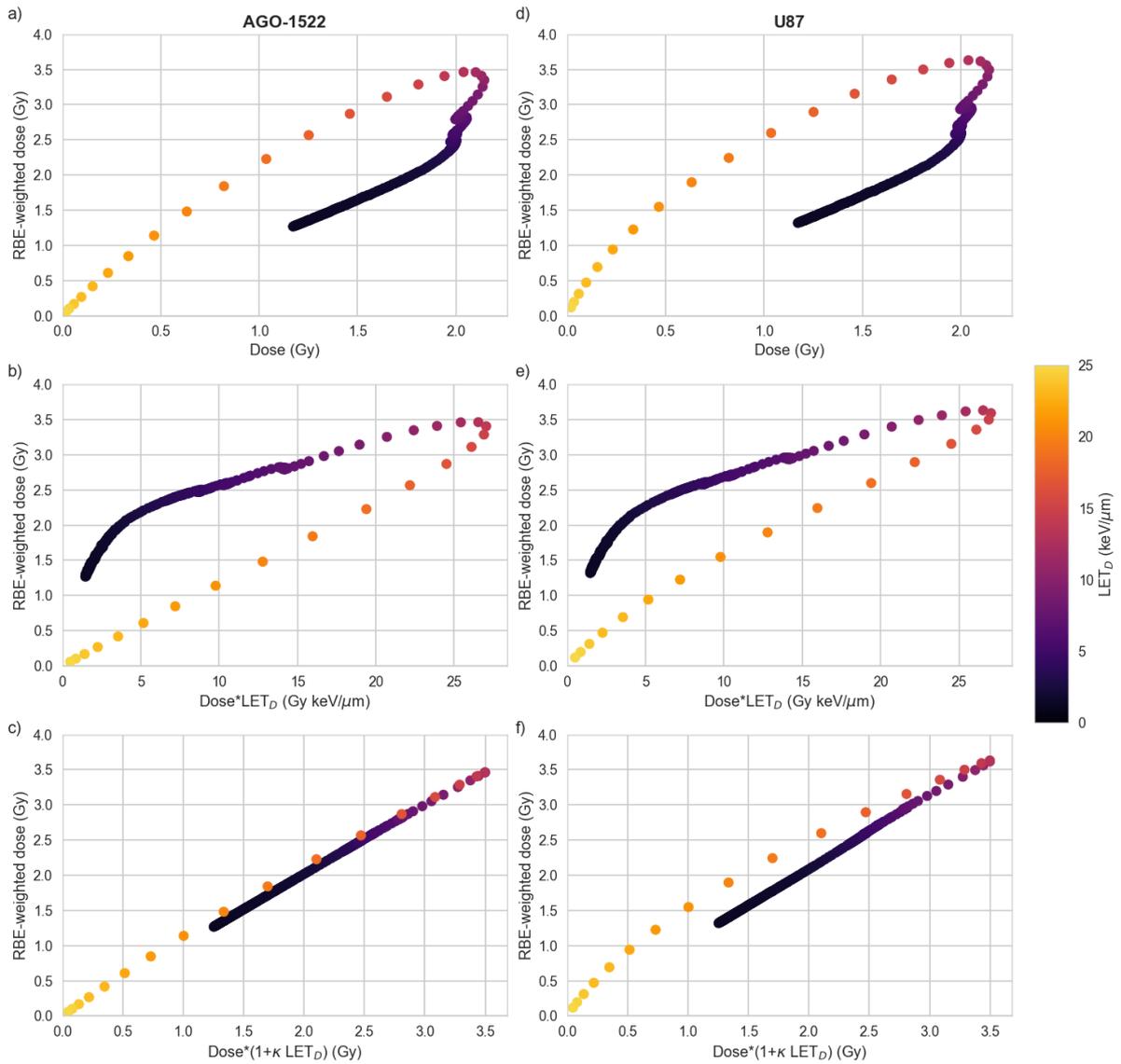

Figure 3: Correlation between RBE-weighted dose and response metrics across a SOBP in AGO-1522 and U87 cells. Each point represents a 0.1 mm slice of the SOBP, coloured according to LET. For dose alone (top), there is a significant biological uncertainty due to variations in LET, under-estimating effects in high LET regions. As in cell survival, this trend is reversed when considering $D \times LET_D$ (middle), but the total biological uncertainty is comparable. However, the weighted-dose metric (bottom) greatly reduces this uncertainty in both cell lines.

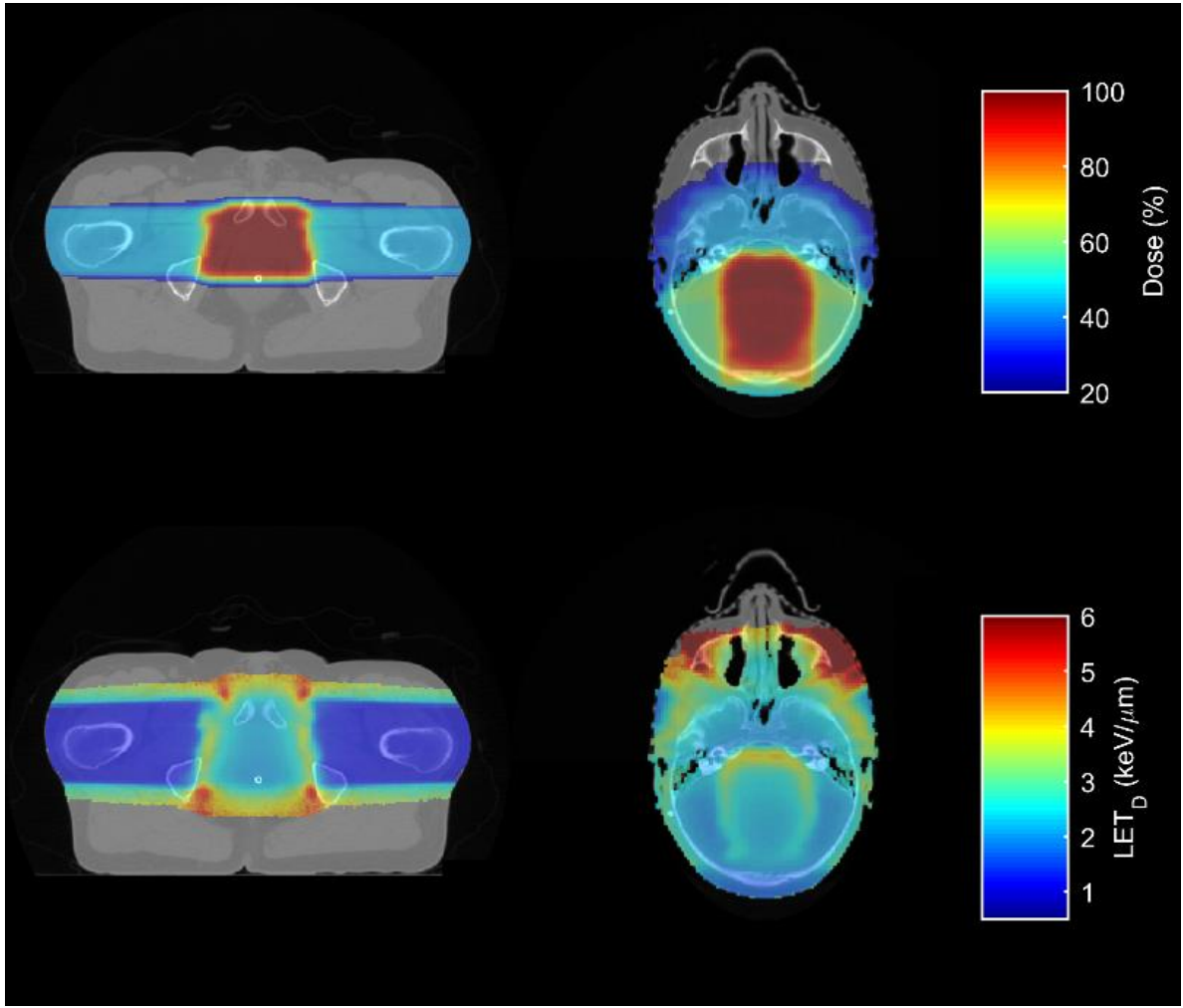

Figure 4: Dose (top) and $LET_D$ (bottom) distributions for prostate (left) and medulloblastoma (right) treatment plans. Both plans deliver a range of doses and LETs delivered using different beam geometries, and have been analysed to explore spatial variations in RBE-weighted dose with other metrics.

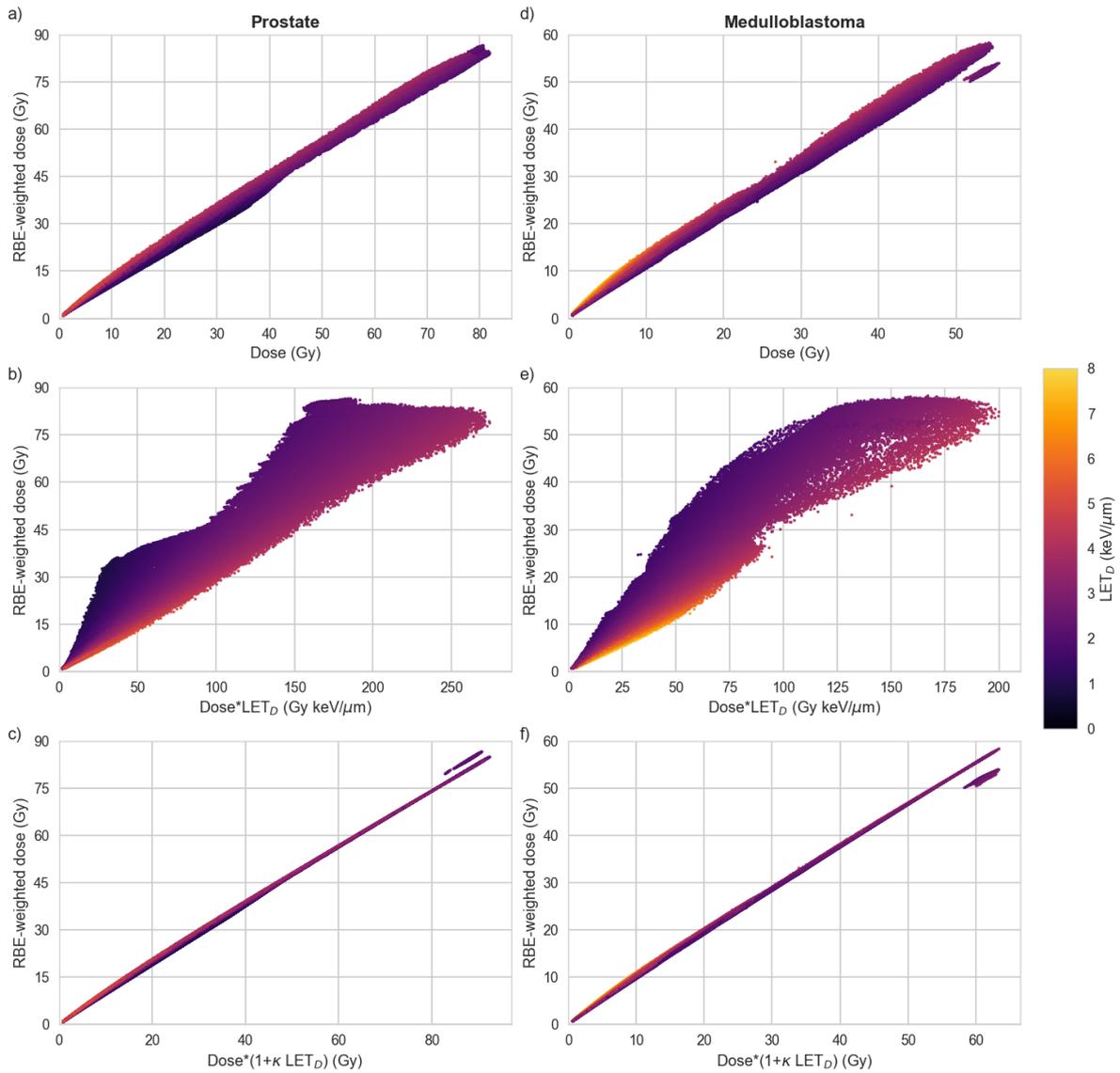

Figure 5: Comparison between RBE-weighted dose and metrics in clinical plans for prostate (left) and medulloblastoma (right). Point clouds have been plotted for each voxel receiving more than 1% of the prescription dose, coloured according to $LET_D$. Dose alone (top) shows the biological variation inherent in clinical plans, with voxels assumed to see the same biological effect showing a variation equivalent to ±5% of the prescription dose, with high $LET_D$ points seeing higher effects. $D \times LET_D$ (middle) reverses this trend, but shows a much broader range of biological uncertainty. CWD (bottom) resolves much of this biological uncertainty, with voxels of a given CWD having a very small spread in RBE-weighted dose.

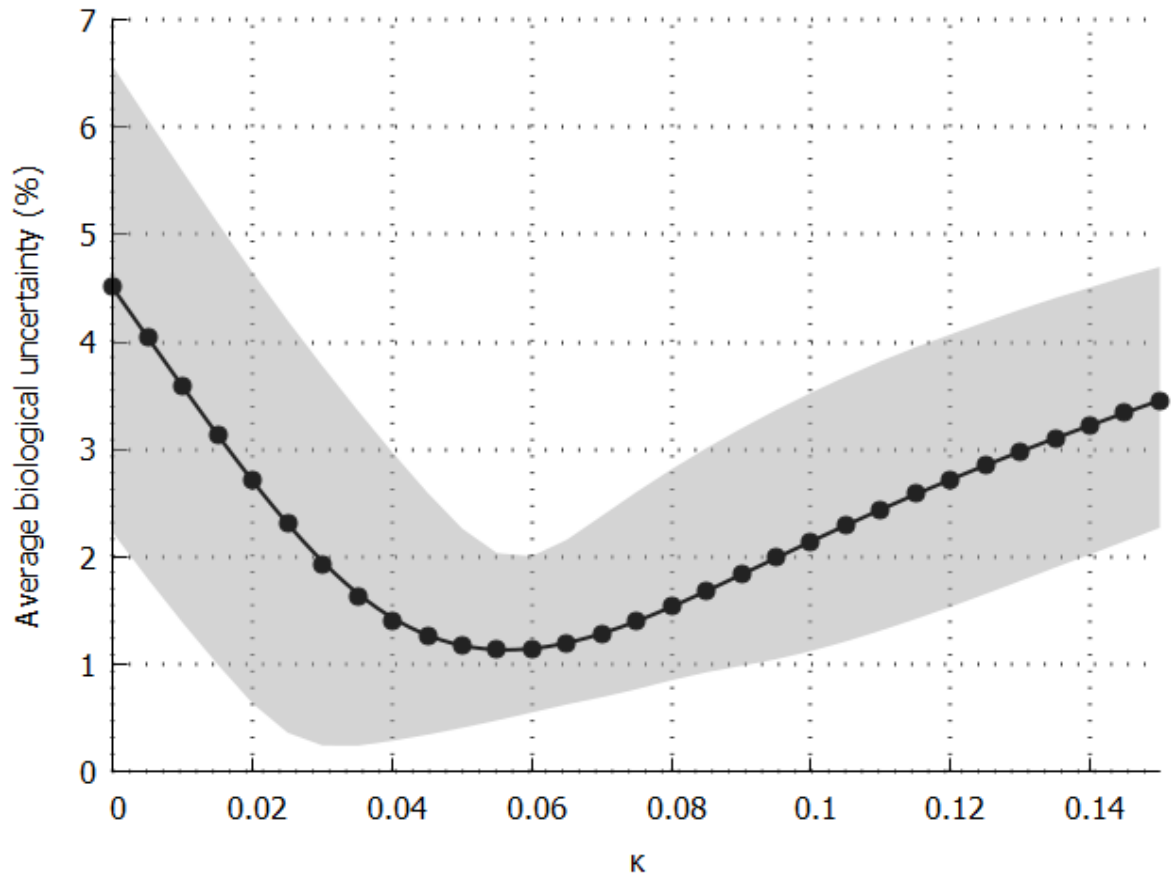

Figure 6: Average biological uncertainty in both treatment plans as a function of κ, for tissues with $\alpha/\beta$ values between 2 and 10 Gy. Points represents average response across all $\alpha/\beta$ values, while shaded region indicates full range. For $\kappa = 0$ this represents the uncertainty associated with using dose alone, and decreases to a minimum of approximately 1% around the $\kappa = 0.055$ used in this work.

# References

[1] Newhauser WD, Zhang R. The physics of proton therapy. Phys Med Biol 2015;60:R155–209. doi:10.1088/0031-9155/60/8/R155.

[2] Paganetti H. Range uncertainties in proton therapy and the role of Monte Carlo simulations. Phys Med Biol 2012;57. doi:10.1088/0031-9155/57/11/R99.

[3] Mohan R, Peeler CR, Guan F, Bronk L, Cao W, Grosshans DR. Radiobiological issues in proton therapy. Acta Oncol (Madr) 2017;56:1367–73. doi:10.1080/0284186X.2017.1348621.

[4] Paganetti H. Relative biological effectiveness (RBE) values for proton beam therapy. Variations as a function of biological endpoint, dose, and linear energy transfer. Phys Med Biol 2014;59:R419-72. doi:10.1088/0031-9155/59/22/R419.

[5] Carabe A, Moteabbed M, Depauw N, Schuemann J, Paganetti H. Range uncertainty in proton therapy due to variable biological effectiveness. Phys Med Biol 2012;57:1159–72. doi:10.1088/0031-9155/57/5/1159.

[6] Carabe-Fernandez A, Dale RG, Jones B. The incorporation of the concept of minimum RBE (RbEmin) into the linear-quadratic model and the potential for improved radiobiological analysis of high-LET treatments. Int J Radiat Biol 2007;83:27–39. doi:10.1080/09553000601087176.

[7] Wedenberg M, Lind BK, Hårdemark B. A model for the relative biological effectiveness of protons: the tissue specific parameter α/β of photons is a predictor for the sensitivity to LET changes. Acta Oncol (Madr) 2013;52:580–8. doi:10.3109/0284186X.2012.705892.

[8] McNamara AL, Schuemann J, Paganetti H. A phenomenological relative biological effectiveness (RBE) model for proton therapy based on all published *in vitro* cell survival data. Phys Med Biol 2015;60:8399–416. doi:10.1088/0031-9155/60/21/8399.

[9] Webb S, Nahum AE. A model for calculating tumour control probability in radiotherapy including the effects of inhomogeneous distributions of dose and clonogenic cell density. Phys Med Biol 1993;38:653–66. doi:10.1088/0031-9155/38/6/001.

[10] Bentzen SM, Christensen JJ, Overgaard J, Overgaard M. Some methodological problems in estimating radiobiological parameters from clinical data-alpha/beta ratios and electron RBE for cutaneous reactions in patients treated with postmastectomy radiotherapy. Acta Oncol (Madr) 1988;27:105–16. doi:10.3109/02841868809090330.

[11] Paganetti H. Relating the proton relative biological effectiveness to tumor control and normal tissue complication probabilities assuming interpatient variability in α/β. Acta Oncol (Madr) 2017;56:1379–86. doi:10.1080/0284186X.2017.1371325.

[12] West CM, Davidson SE, Roberts S a, Hunter RD. Intrinsic radiosensitivity and prediction of patient response to radiotherapy for carcinoma of the cervix. Br J Cancer 1993;68:819–23.

[13] Scheenstra AEH, Rossi MMG, Belderbos JSA, Damen EMF, Lebesque J V., Sonke JJ. Alpha/beta ratio for normal lung tissue as estimated from lung cancer patients treated with stereotactic body and conventionally fractionated radiation therapy. Int J Radiat Oncol Biol Phys 2014;88:224–8. doi:10.1016/j.ijrobp.2013.10.015.

[14] Alsbeih G, Al-Meer RS, Al-Harbi N, Bin Judia S, Al-Buhairi M, Venturina NQ, et al. Gender bias in individual radiosensitivity and the association with genetic polymorphic variations. Radiother Oncol 2016;119:236–43. doi:10.1016/j.radonc.2016.02.034.

[15]  Bassler N, Jäkel O, Søndergaard CS, Petersen JB. Dose-and LET-painting with particle therapy. Acta Oncol (Madr) 2010;49:1170–6. doi:10.3109/0284186X.2010.510640.

[16]  Giantsoudi D, Grassberger C, Craft D, Niemierko A, Trofimov A, Paganetti H. Linear energy transfer-guided optimization in intensity modulated proton therapy: Feasibility study and clinical potential. Int J Radiat Oncol Biol Phys 2013;87:216–22. doi:10.1016/j.ijrobp.2013.05.013.

[17]  Wan Chan Tseung HS, Ma J, Kreofsky CR, Ma DJ, Beltran C. Clinically Applicable Monte Carlo–based Biological Dose Optimization for the Treatment of Head and Neck Cancers With Spot-Scanning Proton Therapy. Int J Radiat Oncol Biol Phys 2016;95:1535–43. doi:10.1016/j.ijrobp.2016.03.041.

[18]  Unkelbach J, Botas P, Giantsoudi D, Gorissen B, Paganetti H. Reoptimization of intensity-modulated proton therapy plans based on linear energy transfer. Int J Radiat Oncol • Biol • Phys 2016;96:1097–106. doi:10.1016/J.IJROBP.2016.08.038.

[19]  Chaudhary P, Marshall TI, Perozziello FM, Manti L, Currell FJ, Hanton F, et al. Relative biological effectiveness variation along monoenergetic and modulated Bragg peaks of a 62-MeV therapeutic proton beam: A preclinical assessment. Int J Radiat Oncol Biol Phys 2014;90:27–35. doi:10.1016/j.ijrobp.2014.05.010.

[20]  Marshall T. II, Chaudhary P, Michaelidesová A, Vachelová J, Davídková M, Vondráček V, et al. Investigating the Implications of a Variable RBE on Proton Dose Fractionation Across a Clinical Pencil Beam Scanned Spread-Out Bragg Peak. Int J Radiat Oncol Biol Phys 2016;95:70–7. doi:10.1016/j.ijrobp.2016.02.029.

[21]  Agostinelli S, Allison J, Amako K, Apostolakis J, Araujo H, Arce P, et al. GEANT4 - A simulation toolkit. Nucl Instruments Methods Phys Res Sect A Accel Spectrometers, Detect Assoc Equip 2003;506:250–303. doi:10.1016/S0168-9002(03)01368-8.

[22]  Perl J, Shin J, Schümann J, Faddegon B, Paganetti H. TOPAS: An innovative proton Monte Carlo platform for research and clinical applications. Med Phys 2012;39:6818–37.

[23]  Underwood T, Paganetti H. Variable Proton Relative Biological Effectiveness: How Do We Move Forward? Int J Radiat Oncol Biol Phys 2016;95:56–8. doi:10.1016/j.ijrobp.2015.10.006.

[24]  Buchsbaum JC, McDonald MW, Johnstone PAS, Hoene T, Mendonca M, Cheng CW, et al. Range modulation in proton therapy planning: A simple method for mitigating effects of increased relative biological effectiveness at the end-of-range of clinical proton beams. Radiat Oncol 2014;9. doi:10.1186/1748-717X-9-2.